# High-Throughput Virtual Screening of 4487 flavonoids: New insights on the structural inhibition of SARS-CoV-2 Main Protease

Gabriel M Jiménez-Avalos[1]§, Ana Paula Vargas-Ruiz[1]§, Nicolás E Delgado-Pease[1]§, Gustavo E Olivos-Ramírez[1], Patricia Sheen-Cortavarría[1], Manolo Fernandez[2], Mirko J Zimic-Peralta[1 2]*, for the COVID UPCH-FARVET Research Group in Peru.

AFFILIATIONS:

1. Laboratorio de Bioinformática y Biología Molecular, Facultad de Ciencias y Filosofía, Departamento de Ciencias Celulares y Moleculares, Universidad Peruana Cayetano Heredia, Lima, Perú.

2. FARVET. Chincha, Perú.

§ These authors contributed equally.

*Corresponding authors: mirko.zimic@upch.pe

**ABSTRACT**

COVID-19 presents a great threat to public health worldwide and the infectious agent SARS-CoV-2 is currently the target of much research aiming at inhibition. The virus' main protease is a dimeric enzyme that has only recently begun to be thoroughly described, opening the door for virtual screening more broadly. Here, a PAIN-filtered flavonoid database was screened against four sites of the protease: a free (normal) conformation of the Substrate Binding Site (NSBS), an induced-fit state of the SBS (ISBS), a Dimerization Site (DS) and a Cryptic Site (CS). The mean binding energy of the top five ligands from each site were -9.52, -11.512, -7.042 and -10.348 kcal/mol for the NSBS, the ISBS, the DS and the CS, respectively. For the DS and CS, these top five compounds were selected as candidates to bind their respective site. In the case of SBS, the top 30 ligands with the lowest binding energies from NSBS and ISBS were contrasted and the ones present in both lists were selected as the final candidates. The final list was: Dorsilurin E (FL3FQUNP0001), Euchrenone a11 (FL2FALNP0014), Kurziflavolactone C (FL2FA9NC0016), Licorice glycoside E (FL2F1AGSN001) and Taxifolin 3'- (6"-phenyl- acetylglucoside) (FL4DACGS0020) for the SBS; Sanggenol O (FL2FALNP0020), CHEMBL2171573, Kanzonol E (FL3F1ANP0001), CHEMBL2171584 and Abyssynoflavanone VI (FL2FACNP0014) for DS and CHEMBL2171598, CHEMBL2171577, Denticulaflavanol (FL5FAANR0001), Kurzichalcolactone (FL1CA9NC0001) and CHEMBL2171578 for CS. Virtual screening integrated several confirmation methods, including cross-docking assays and positive and negative controls. All 15 compounds are currently subjected to molecular dynamics so as to theoretically validate their binding to the protease.

## 1. INTRODUCTION

The Coronavirus Disease-2019 (or COVID-19) was first reported in late December 2019 in Wuhan, PRC (1). First identified as a kind of pneumonia, common symptoms are fever, cough and myalgia or fatigue, with less common symptoms ranging from sputum production, headache, haemoptysis and diarrhoea (2). Later developments included lymphopenia and dyspnoea. These findings point at a clinical description akin to that of severe acute respiratory syndrome (SARS) coronavirus (2,3).

Nonetheless, finding a treatment for viral diseases seldom is a serendipitous event. It requires systematic narrowing of candidates, usually by virtual screening (VS) methods. VS has been able to provide ample and robust support for drug discovery, design and evaluation efforts (4,5) with a variety of specialized options (6) as the *in silico* counterpart for high throughput screening. These are often coupled in sequential or parallel arrays: hierarchical VS (HLVS) or parallel VS (PVS), respectively (5). This development was inevitable given both the relentless growth in library sizes (7,8) and the equally vertiginous development in the virtual screening *par excellence*, molecular docking (9–11). HLVS has been invaluable in sieving through large databases of small organic molecules to find drugs for viral diseases. This is invaluable for the development of new contenders as the current pharmacological options are rendered obsolete by high viral genome variability (12). Such developments of new drugs against virus through VS include influenza (13), ebola (14) and dengue (15).

Trying SARS-CoV drugs against SARS-CoV-2 has been proposed (16) given the close genomic proximity between them (17). Unfortunately, those drugs have not been universally approved for treatment and usually remain stagnant on clinical or pre-clinical trials (18). Moreover, even small differences in amino acid sequence can affect drug efficaciousness (19). Unsurprisingly, the appeal of VS approaches against SARS-CoV-2 has not eluded scientists during this pandemic. Most have tried to evaluate the repurposing of existing drugs (16,20–22). Nevertheless, these research lines run the risk of facing adverse effects later on for employing drugs designed for other sites (23). Other papers have focused on the active site of SARS-CoV-2 main protease ($M^{PRO}$) (16,24), but this emphasis could be too restrictive since dimerization inhibition is a plausible alternative (25,26). Furthermore, the occurrence of cryptic sites is a possibility worth exploring. Loosely put, a binding site can be described as cryptic if identifiable in the ligand-bound while not in the unbound structure of a given protein (27). These locations pose a promising way to increase the druggability of a target protein (28,29). Therefore, an exhaustive VS of readily available, pharmacologically safe, new small molecules targeted at more than one site in the $M^{PRO}$ is the aim of this paper.

Flavonoids are one such group of small molecules produced as secondary metabolites in fungi and plants. They present a common diphenyl structure coupled with a heterocyclic ring and with various hydroxyl phenolic groups with great chelation capacity for iron and other transition metals. Especially relevant to this research is the antiviral activity of flavonoids. The flavonoid apigenin combined with acyclovir has a significant effect on types 1 and 2 herpes virus simplex known since the early 1990s (30). Remarkably, 64 flavonoids were previously screened against the SARS-CoV-1 3CL (chymotrypsin-like) protease (3CLpro) with promising results (31). Furthermore, the 3CLpro from SARS-CoV-1 differs from the SARS-CoV-2 $M^{PRO}$ in but 12 amino acids, which justifies a further attempt for flavonoid inhibition against this second protein (32). In that vein, a handful of flavonoids have been recently tried *in silico* against SARS-CoV-2 spike protein (33), which leaves open the questions as to what other flavonoids could prove useful. All this considered, the panoply of flavonoids available in many online databases presents itself as an ideal opportunity to fuel a new quest for SARS-CoV-2 inhibition.

## 2. METHODS

### 2.1. Generation of 3D structures and selection of putative binding sites

#### 2.1.1. Library preparation

An exhaustive search was conducted for compounds reported as flavonoids in the Metabolomics (34) database with an inhouse R script for web scraping using the rvest package (35). Similarly, the online databases DrugBank (36), CHEMBL (37) and PubChem (38) were subject to an automated search. For these last websites, searches were performed manually with the key words flavonoid, flavonoids and flavo, except for DrugBank, where the whole base was downloaded. Then, all four bases were standardized to show a unique ID, SMILE and online source for each compound. All four databases from DrugBank, PubChem, CHEMBL and Metabolomics were merged through an R script, additionally parsing the complete database through the same means in order to remove compounds with molar weight under 180 kDa. Compounds with identical SMILES were removed for each database.

#### 2.1.2. PAINs filtering

The database was then uploaded to FAFDrugs4 server (39) for further processing. Two SMILES, even though literally distinct, may be synonyms for the same compound. FAFDrugs4 reads each SMILE and interprets the code into structure in order to exclude synonymic SMILEs. Next, all screening tools to identify and purge pan-assay interference compounds (PAINS) offered by the website were employed. A final SMILE database was downloaded and matched for their unique IDs and online sources.

#### 2.1.3. 3D structure generation.

SMILEs were read and employed to generate .smi files for each compound. Afterwards, these files worked as blueprints for the generation of .sdf files containing the bidimensional structure of each compound through the package OpenBabel v.3.1.1 (40). These in turn were the starting point for the three-dimensional structures, each of which was subject to a 25000-step minimization through the steepest-descent method (41). Finally, the resulting .sdf files of the minimized compounds were converted to .pdb format for their docking preparation. All these steps were performed automatically through an inhouse Python script.

#### 2.1.4. Electrochemical and structural assessment of putative binding sites

Three sites of interest were identified for docking targeting: the substrate-binding site (SBS), the dimerization site (DS) and the cryptic site (CS). The first two were chosen after bibliographic consideration (32,42) while the cryptic site was originally predicted using CryptoSite (28). Collectively, they are referred to as "putative binding sites".

pH is not homogeneously distributed along the protein surface, given that pH in each region is determined by the microenvironment fostered by the residues specific to that region. In order to analyse these microenvironments, server PDB2PQR (43) was employed. Said program assigns the appropriate protonation states to an input protein's residues, presently the apo-structure of $M^{PRO}$ excised from PDB ID: 6LU7. This software takes into account the surrounding residues as well as the macro-system pH, which in this case was set to 7.4 (physiological pH). Thus, the protonation states of key histidines were employed to infer the microenvironment pHs for the three putative binding sites and where they stood in comparison to the general pH of the system.

Each putative binding site had its respective solvent-accessible surface area (SASA) found in PyMol v.2.4 (44), using as input the hydrogen-curated structure of $M^{PRO}$ obtained before. All the regions were scoured for residues capable of establishing strong interactions, aromatic rings and other relevant moieties. The exact residues which conform the S1' and S1-S5 subsites of the SBS region (42,45) were determined by analysis of $M^{PRO}$-N3 complex (PDB: 6LU7).

## 2.2. Docking assays

### 2.2.1. Ligand preparation.

20 random compounds were selected in order to conclude the structure was correctly represented. Once this was settled, the files were converted from .pdb to .pdbqt format with the AutodocksTool v.4.2 package (46). All torsions were assigned and Gasteiger charges were added for all atoms in the compound structures.

### 2.2.2. Receptor preparation.

$M^{PRO}$ structure was procured in two distinct conformations: a free conformation (PDB: 6YB7) and another conformation induced by an inhibiting ligand in the active site (PDB: 6LU7). These structures were independently curated on PDBfixer (47) with the intent of repairing lost atoms and non-standard residues, as well as for deleting heteroatoms. Using this repaired chains, both receptors were constructed as dimers from the available protomeric states in their respective .pdb files (biological assembly) using PyMol software.Then, all receptor structures were converted to .pdbqt format using the graphic interface found in the AutodockTools v.4.2 package, adding polar hydrogens and Gasteiger charges as well.

### 2.2.3. Virtual Screening.

For the following procedure, a semiflexible VS (with rigid receptor and flexible ligand) was conducted so as to assess the possibility of compound interaction with each putative binding site. For this purpose, Autodock Vina v.1.1.2 (48) software was employed with 24 of exhaustiveness (49) through a Python script which automated the task.

#### 2.2.3.1. SBS

In order to reach robust conclusions for SBS-binding compounds, a cross-docking methodology was followed. To this end, the receptors were free $M^{PRO}$ (PDB: 6YB7) and ligand-induced conformation $M^{PRO}$ (PDB: 6LU7). All residues conforming the SBS were included within the volume of the search box. Box coordinates for normal and induced-fit docking were 12.339, 1.287, 23.152 and -15.117, 14.564, 67.870; respectively. Dimensions for both boxes were kept at 35 Å x 35 Å x 35 Å for minimizing search bias.

#### 2.2.3.2. DS and CS

Given that no induced conformations were available for DS or CS, no cross-docking was performed, keeping free $M^{PRO}$ as the receptor. Box coordinates for DS and CS were 1.738, -3.380, 4.457 (22 Å × 28 Å × 36 Å) and 9.104, 12.126, -6.685 (30 Å × 40 Å × 30 Å); respectively.

### 2.3. Validation

#### 2.3.1. Validation docking

In order to confirm VS results, the next step was a validation docking (VD) with more exhaustive parameters using the top 100 best compounds in Autodock-GPU v.4.2 (50). Lamarckian genetic algorithm (LGA) was used as a global search method, while the local search was directed after the Solis-Wets algorithm (SWA). The procedure included 25 million evaluations and 150 runs. Box coordinates were maintained irrespective of each virtual screening, with 0.375 Å spacing. Dimensions were 18.75 Å × 18.75 Å × 18.75 Å, 22.5 Å × 30 Å × 37.5 Å, and 30.375 Å × 40.5 Å × 30.375 Å for the SBS (both NSBS and ISBS), DS and CS VDs, respectively.

For each docking assay, the best pose was selected based on its 2Dscore (51). Briefly, each pose is assigned a score based on the standard normalization (Zscore) of two variables: binding energy (ΔG) and cluster population (Pop). Said score is calculated as follows:

$$2D_{score} = -1 \times [Z_{score}(\Delta G)] + Z_{score}(Pop)$$

The selected solutions were then ranked according to their binding energies. For the DS and CS VDs, the top five ligands with the lowest binding energies were selected as hits. In the case of SBS, the top 30 ligands with the lowest binding energies from the VDs against NSBS and ISBS were contrasted and the ones present in both lists were selected as hits. Then, their most favourable pose was extracted using PyMol, converted to .mol2 format and supplemented with the missing hydrogen atoms using OpenBabel at pH 7.4. The protein-ligand complex was reconstructed by merging the ligand file with its corresponding hydrogen-curated receptor (see Section 1.3 above).

#### 2.3.2. Positive and negative controls

It is key to confirm that the docking workflow can replicate the binding event of a known experimental inhibitor and provide reference values (positive control). To this end, the potent broad-spectrum non-covalent SARS-CoV-2 inhibitor X77 was chosen as comparative standard. The ligand (X77) was excised from a X77-$M^{PRO}$ co-crystallized complex (PDB: 6W63) using PyMol, while the same was done for the receptor (free $M^{PRO}$) from a N3-$M^{PRO}$ co-crystallized complex (PDB: 6LU7). Then, the ligand .pdb file was converted into .sdf using OpenBabel v.3.1.1. This file was curated manually using PyMol again so as to correct bond order and add missing hydrogen atoms. X77 was then docked against ISBS in order to produce a cross-docking while maintaining the last VD's parameters. The resulting complex was contrasted with the initial structure (PDB: 6W63) by their RMSD using PyMol (52).

Equally important is the capacity of sieving protocols for rejecting inert molecules. These compounds could exhibit false positives in biochemical assays due to colloidal aggregation (53,54). This is the case of TDZD-8, which only inhibits $M^{PRO}$ through aggregation (45). Thus, this molecule is taken as an appropriate negative control to test the efficacy of the present protocol at excluding unspecific, promiscuous molecules. Again, the docking of this compound follows the VD protocol explained before.

### 2.4. Protein-ligand interaction analysis

Protein–Ligand Interaction Profiler (PLIP) was used with default parameters to further explore protein-ligand interactions arising from docking (55). Coupled with previous electrochemical-structural revision, PLIP shed light on the chemical environment surrounding each flavonoid.

Regarding the SBS, it is worth mentioning that only the interactions provided by the VD against ISBS were identified and analysed.

In addition to these steps, the protein-ligand complex .pdb files for each hit (see Section 2.3.1 above) was uploaded to the online service Dr. SASA (56) in order to obtain surface data.

# 3. RESULTS

## 3.1. Flavonoid Structural Database

The complete database (before FAFDrug4 filters) included 6696 compounds. The final database comprised 4887 compounds. SMILE information for every compound sufficed to generate satisfactory minimized 3-D structures in .pdbqt format. The approximate mean torsion quantity was 9.73. Exhaustiveness was adjusted to 24 for Vina VS and the number of evaluations was set to 25 million for the VD as a higher number of torsions requires deeper search standards.

## 3.2. Cryptic Site Description

Cryptic site prediction was done through CryptoSite, which produces a .pol.pred file containing prediction scores for every residue. The residue is presumed to have a high probability of constituting a cryptic site whenever the score is above 10. Out of these, several were also present in the same region as the SBS or DS boxes or relatively close to their residues. These were left out for the sake of dynamic independence. The residues comprising CS are Lys5, Met6, Pro108, Gly109, Arg131, Trp218, Phe219, Tyr239, Glu240, Leu271, Leu272, Leu287, Glu288, Asp289, Glu290, Arg298, Gln299 and Val303.

### 3.3. Analysis of Putative Binding Sites

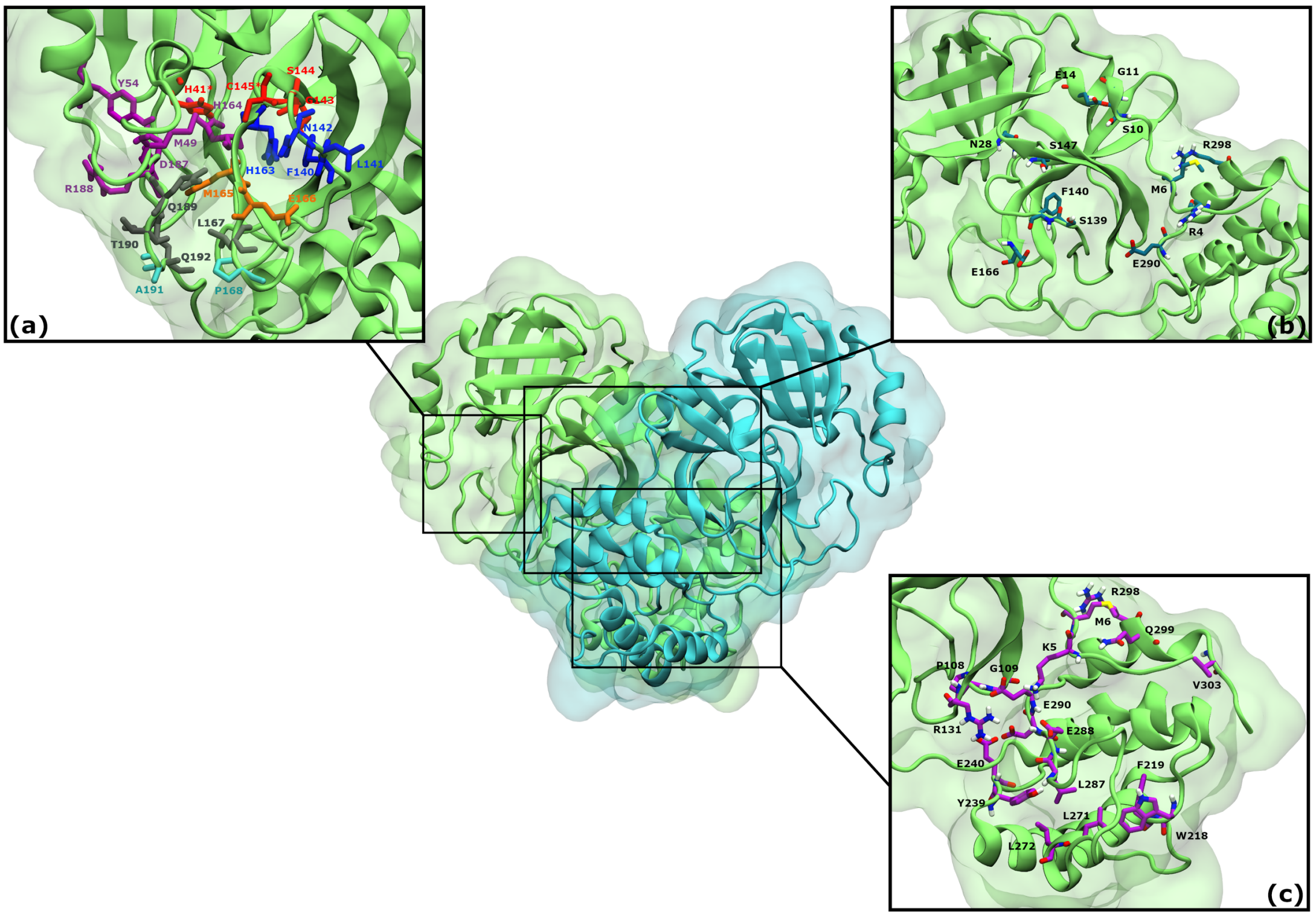

**Figure 1. The M$^{PRO}$ of SARS-CoV-2 and its putative binding sites.** Monomers are in cartoon and surface representations, each one in a different colour. The region covered by each putative binding site in the green monomer is shown as a black rectangle. Residues conforming the sites are depicted as sticks along their names. (a) Substrate binding site, with colours representing subregions: red for S1', blue for S1, purple for S2, orange for S3, grey for S4 and cyan for S5. Catalytic residues are marked with an asterisk. (b) Dimerization site. (c) Cryptic site.

M$^{PRO}$ is a homodimer with a heart-shaped structure (Figure 1). The SBS region of M$^{PRO}$ can be further subdivided into S1' and S1-S5 subsites. Respectively, these sites interact with P1' and P1-P5 segments of a generic substrate peptide (42). S1' is defined to be the catalytic region. Other segments of the peptide are numbered consecutively, starting from P1'. According to this definition, it is possible to map S1' and S1-S5 subsites by analysing the M$^{PRO}$-N3 interaction described elsewhere (45). Shortly, the S1' subsite comprises His41, Gly143, Ser144 and Cys145; S1 includes Ser1 (from the other protomer), Phe140, Leu141, Asn142 and His163; S2 consists in Met49, Tyr54, His164, Asp187 and Arg188; S3 consists in Met165 and Glu166; S4 is constituted by Leu167, Gln189, Thr190 and Gln192; and S5 encompasses Pro168 and Ala191 (Figure 1a). All of these residues are part of, but not limited to the ones reported in the literature (32). Regarding dimensions, SBS has a SASA of 975.013 Å$^2$. Two histidine tautomers may exist under physiological conditions: HSD (protonated in the delta position) with a pKa of 6.7, and HSE (protonated in the epsilon position) with a pKa of 7.3 (57). In the SBS, three monoprotonated histidines were found: His41 and His163 (HSD) and His164 (HSE). Apart from histidines, three amino acids with charged side chains at physiological pH were observed within the site (Table 1). Lastly, Cys145 forms a catalytic dyad with His41.

| Site | Residue | Charged | Polar |
|---|---|---|---|
| Substrate | Ser1* | 0 | Y |
| | His41 | 0 | Y |
| | Met49 | 0 | N |
| | Tyr54 | 0 | N |
| | Phe140 | 0 | N |
| | Leu141 | 0 | N |
| | Asn142 | 0 | Y |
| | Gly143 | 0 | N |
| | Ser144 | 0 | Y |
| | Cys145 | 0 | N |
| | His163 | 0 | Y |
| | His164 | 0 | Y |
| | Met165 | 0 | N |
| | Glu166 | - | Y |
| | Leu167 | 0 | N |
| | Pro168 | 0 | N |
| | Asp187 | - | Y |
| | Arg188 | + | Y |
| | Gln189 | 0 | Y |
| | Thr190 | 0 | Y |
| | Ala191 | 0 | N |
| | Gln192 | 0 | Y |
| Dimer | Arg4 | + | Y |
| | Met6 | 0 | N |
| | Ser10 | 0 | Y |
| | Gly11 | 0 | N |
| | Glu14 | - | Y |
| | Asn28 | 0 | Y |
| | Ser139 | 0 | Y |
| | Phe140 | 0 | N |
| | Ser147 | 0 | Y |
| | Glu166 | - | Y |
| | Glu290 | - | Y |
| | Arg298 | + | Y |
| Cryptic | Lys5 | + | Y |
| | Met6 | 0 | N |
| | Pro108 | 0 | N |
| | Gly109 | 0 | N |
| | Arg131 | + | Y |
| | Trp218 | 0 | N |
| | Phe219 | 0 | N |
| | Tyr239 | 0 | N |
| | Glu240 | - | Y |
| | Lys271 | + | Y |
| | Lys272 | + | Y |
| | Lys287 | + | Y |
| | Glu288 | - | Y |
| | Asp289 | - | Y |
| | Glu290 | - | Y |
| | Gln299 | 0 | N |
| | Val303 | 0 | N |

**Table 1. Physical-chemical characteristics of residues conforming the putative binding sites.** A list of all the residues comprising putative binding sites is provided. Residues are categorized based on their charge (at physiological pH) and polarity. +: positive charged side chain. -: negative charged side chain. 0: neutral side chain. Y: yes. N: no.

The following residues constitute the DS with a joint SASA of 736.627 Å$^2$: Arg4, Met6, Ser10, Gly11, Glu14, Asn28, Ser139, Phe140, Ser147, Glu166, Glu290 and Arg298 (Figure 1b) (32). Nine out of 12 residues are polar, which raises the chances of forming hydrogen bonds with a ligand. There are no histidines among the constituting residues, yet His172 and His163 (both as HSD) are found near Ser139, Phe140 and Glu166. It is worth noting that the source material that predicted SARS-CoV-2 DS inferred the location of the site based on SARS-CoV-1 DS (32), therefore their site description is perfectible by more direct studies. Similar to the SBS, five amino acids with charged side chains at physiological pH were observed (Table 1).

Finally, the CS residues (Figure 1c) jointly make up a SASA of 665.350 Å2. Unlike DS, only half of the residues were polar. No histidines were observed inside nor in the near vicinity. Nevertheless, this site is the most likely to host saline bonds given the nine residues with ionizable sidechains found within (Table 1).

### 3.4. Virtual Screening

The best 100 compounds for each putative binding site had mean binding energies of -8.89, -10.06, -8.08 and -9.21 kcal/mol for the NSBS, the ISBS, the DS and the CS, respectively. Each of these ligands' sets were selected for a VD against their respective putative binding site.

### 3.5. Validation Docking

Mean binding energies for each site's top fivr were -9.52, -11.512, -7.042 and -10.348 kcal/mol for the NSBS, the ISBS, the DS and the CS, respectively. After cross-referencing the NSBS and ISBS top 30, a total of fifteen compounds were retrieved (Table 2). The final VD results presented the ligands Dorsilurin E (FL3FQUNP0001), Euchrenone a11 (FL2FALNP0014), Kurziflavolactone C (FL2FA9NC0016), Licorice glycoside E (FL2F1AGSN001) and Taxifolin 3'- (6"-phenyl-acetylglucoside) (FL4DACGS0020) binding only the SBS; Sanggenol O (FL2FALNP0020), CHEMBL2171573, Kanzonol E (FL3F1ANP0001), CHEMBL2171584 and Abyssynoflavanone VI (FL2FACNP0014) binding only the DS; CHEMBL2171598, CHEMBL2171577, Denticulaflavanol (FL5FAANR0001), Kurzichalcolactone (FL1CA9NC0001) and CHEMBL2171578 binding only CS (Figure 2). Of these compounds, seven flavonoids (Licorice glycoside E, Taxifolin 3'- (6"-phenylacetylglucoside), CHEMBL2171573, CHEMBL2171584, CHEMBL2171598, CHEMBL2171577 and CHEMBL2171578) had more than 12 torsions, while the rest had less than eight torsions.

**Table 2. List and 2D structure of top flavonoid hits.**

| ID | Common name | Site | SMILE | 2D-Structure |
|---|---|---|---|---|
| **FL2FACNP0014** | Abyssinoflavanone VI | DS | c(c5O)c(cc(c45)OC(CC(=O)4)c(c2)c(C1)c(c(O3)c2CCC(C)(C)3)OC(C1O)(C)C)O | 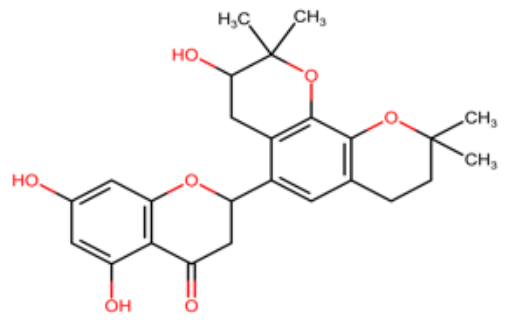 |
| **FL3F1ANP0001** | Kanzonol E | DS | c(c4)(CC=C(C)C)c(cc(c43)OC(=CC3=O)c(c1)cc(C=2)c(OC(C2)(C)C)c1)O | 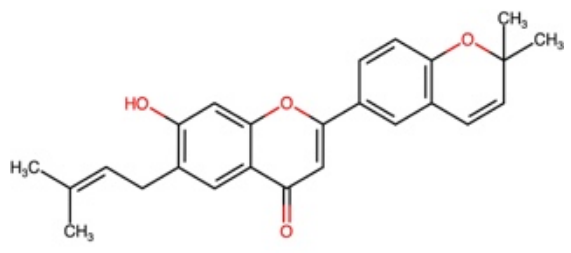 |
| **FL2F1AGSN001** | Licorice glycoside E | SBS | c(OC(O4)C(OC(O5)C(O)C(COC(=O)c(c76)cnc6cccc7)(C5)O)C(C(C4CO)O)O)(c1)ccc(C(O2)CC(c(c3)c2cc(c3)O)=O)c1 | 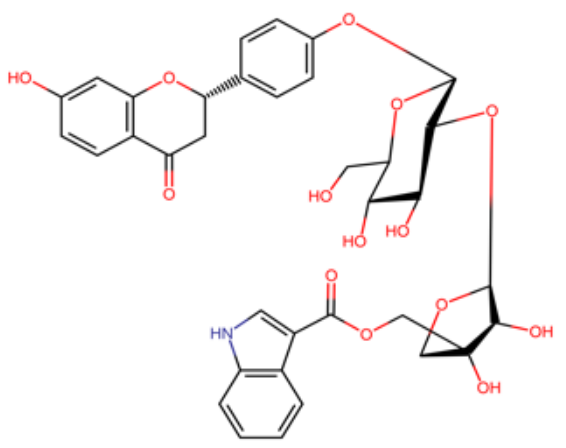 |
| **FL2FALNP0014** | Euchrenone a11 | SBS | O=C(c31)CC(c(c5O)cc(c4c5)C=CC(O4)(C)C)Oc1c(CC=C(C)C)c(c2c3O)OC(C=C2)(C)C | 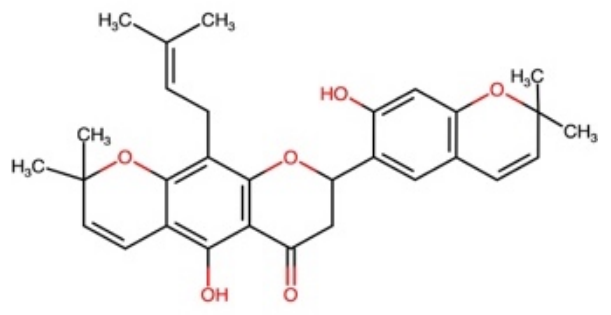 |
| **FL2FALNP0020** | Sanggenol O | DS | c(C(C4)Oc(c5)c(c(cc5O)O)C4=O)(c23)cc(c1c(C=CC(C)(C)O3)2)C=CC(C)(C)O1 | 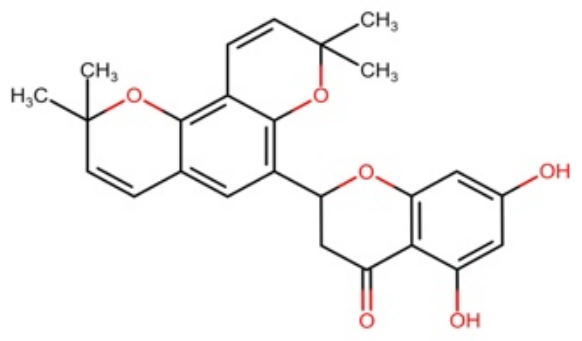 |
| **FL3FQUNP0001** | Dorsilurin E | SBS | c(c54)(O6)c(CCC6(C)C)c(c2c4OC(CC5)(C)C)OC(=c(c3)c(=O)cc(O)c3)C(=C21)CCC(C)(C)O1 | 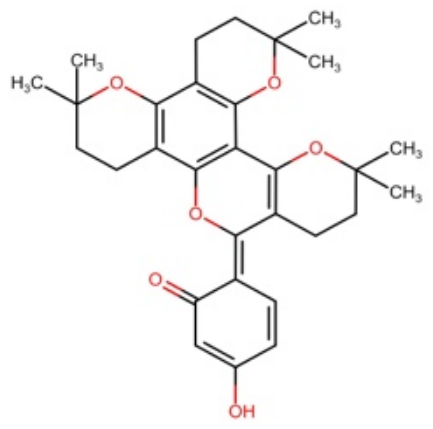 |
| **CHEMBL2171573** | NA | DS | Cc1ccc(COc2cc(OCc3ccc(C)cc3)cc(-c3cc(=O)c4ccc(OCC(O)CNC(C)(C)C)cc4o3)c2)cc1 | 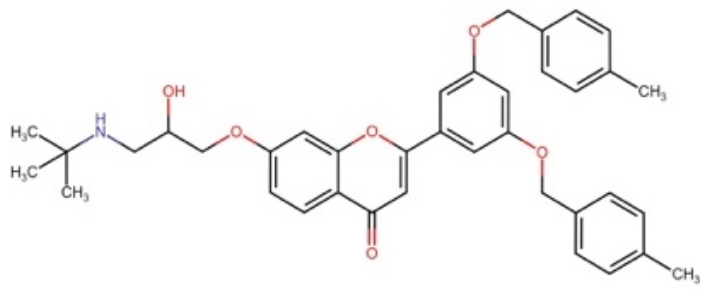 |
| **CHEMBL2171577** | NA | CS | CC(C)(C)NCC(O)COc1ccc2c(=O)cc(-c3cc(OCc4cc(Cl)cc(Cl)c4)cc(OCc4cc(Cl)cc(Cl)c4)c3)oc2c1 | 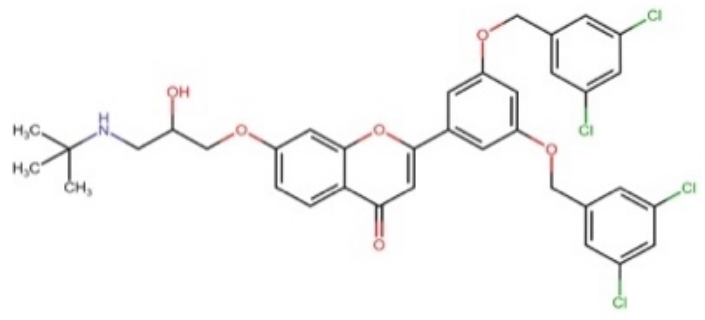 |
| **CHEMBL2171598** | NA | CS | O=c1cc(-c2cc(OCc3ccccc3)cc(OCc3ccccc3)c2)oc2cc(OCC(O)CN3CCN(c4ccccc4)CC3)ccc12 | 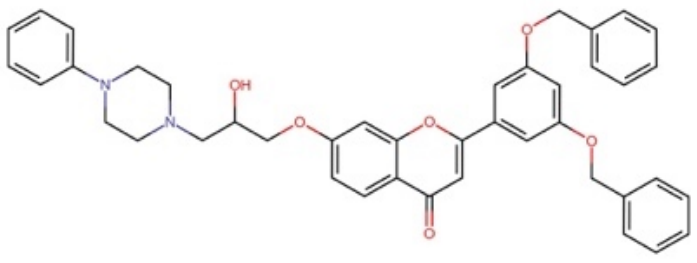 |
| **FL1CA9NC0001** | Kurzichalcolactone | CS | c(c5)ccc(c5)C=CC(C3)(OC(C4)CCCC(Oc(c2C34)cc(c(c(O)2)C(C=Cc(c1)cccc1)=O)O)=O)O | 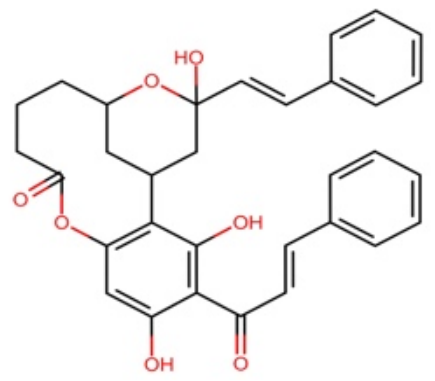 |
| **FL2FA9NC0016** | Kurziflavolactone C | SBS | C(C3=O)CCC(O1)CC(c(c4O)c(cc(O5)c(C(=O)CC(c(c6)cccc6)5)4)O3)CC1(C=Cc(c2)cccc2)O | 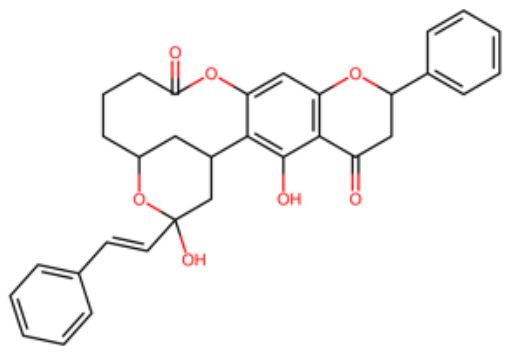 |
| **CHEMBL2171578** | NA | CS | CC(C)NCC(O)COc1ccc2c(=O)cc(-c3cc(OCc4cc(Cl)cc(Cl)c4)cc(OCc4cc(Cl)cc(Cl)c4)c3)oc2c1 | 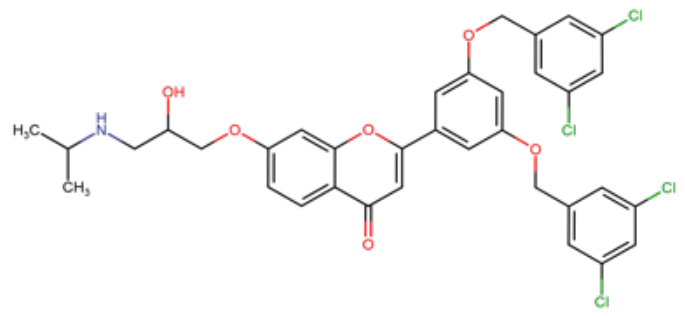 |
| **FL5FAANR0001** | Denticulaflavonol | CS | c(C(O2)=C(C(c(c(O)3)c2cc(c3CC=C(C)CCC(C5(C)4)C(=C)CCC4C(CCC5)(C)C)O)=O)O)(c1)ccc(c1)O | 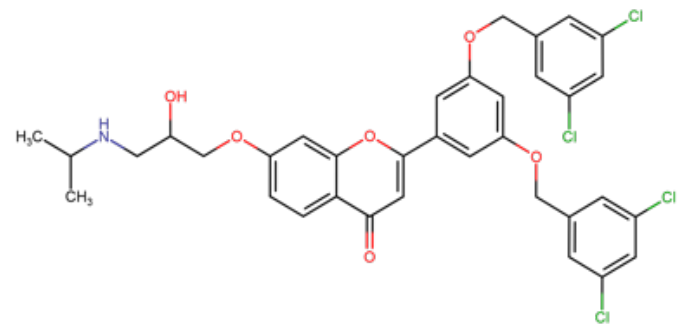 |
| **CHEMBL2171584** | NA | DS | CC(C)(C)NC[C@H](O)COc1ccc2c(=O)cc(-c3cc(OCc4ccccc4)cc(OCc4ccccc4)c3)oc2c1 | 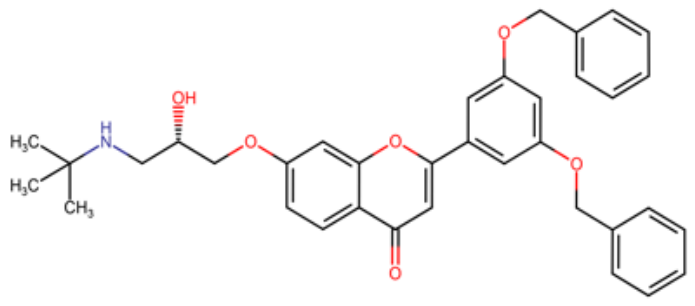 |
| **FL4DACGS0020** | Taxifolin 3'- (6"-phenylacetylglucoside) | SBS | c(c1)(O)cc(c(C(=O)2)c1OC(c(c3)ccc(O)c(OC(O4)C(O)C(C(C(COC(Cc(c5)cccc5)=O)4)O)O)3)C2O)O | 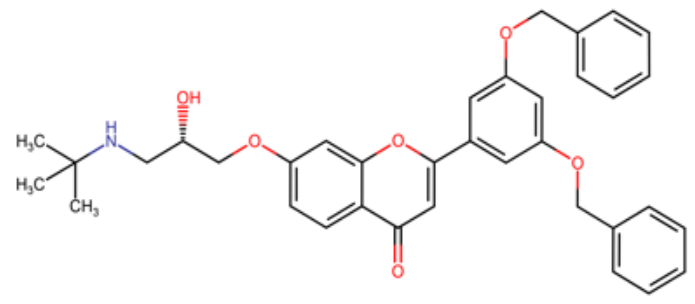 |

The common name, code, SMILE, 2D structure of each flavonoid is indicated, along with the putative binding site it binds to.

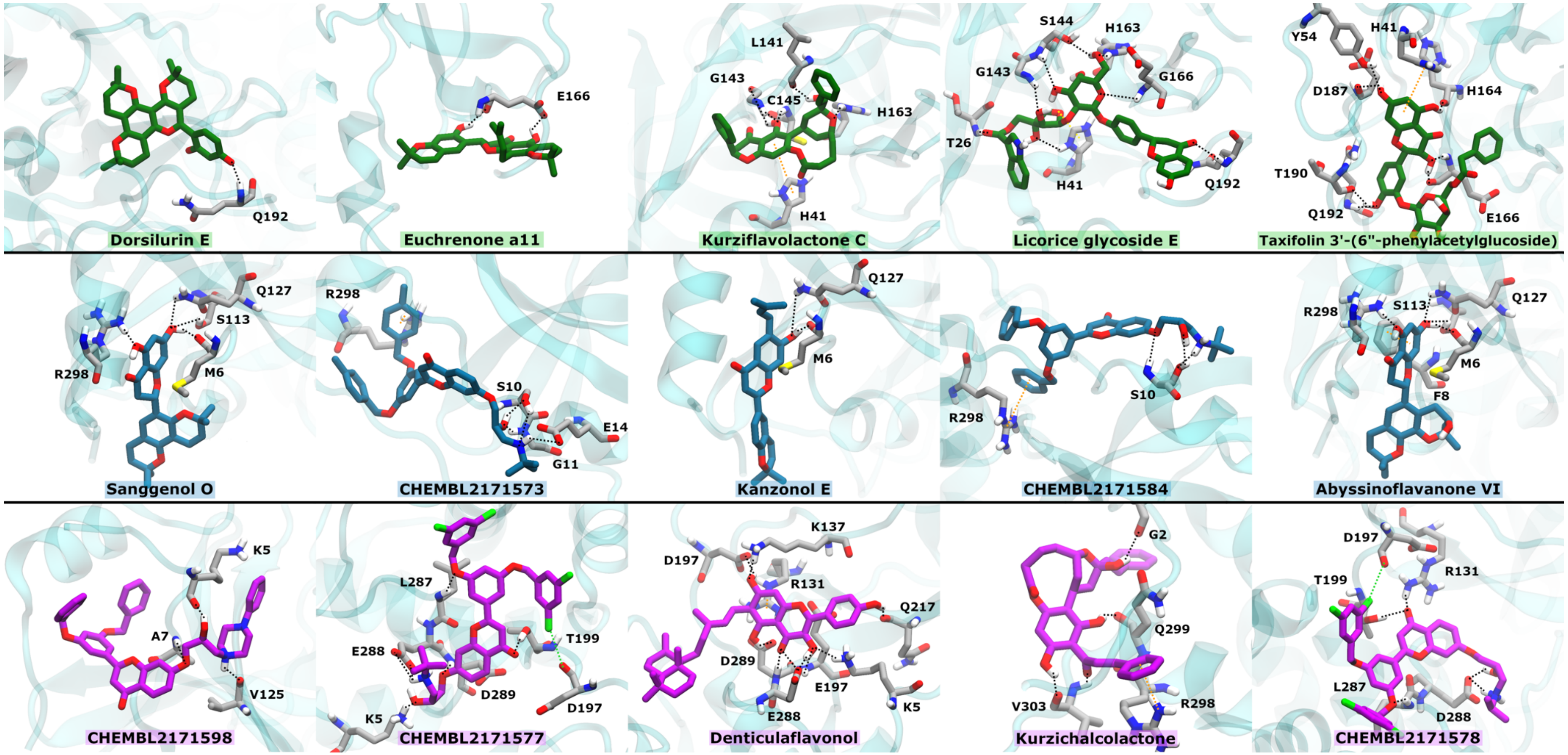

**Figure 2. Interactions of the top 5 binding energy compounds.** Putative binding sites are in order of appearance and the ligands are color-coded: induced-fit substrate binding site (1° row, green), dimerization site (2° row, blue) and cryptic site (3° row, magenta). Ligands are depicted as sticks and protein as cartoon and surface. The common name of each ligand is shown, except for those where only a code is available. Main interactions are shown as dotted points: hydrogen bonds in black, $\pi$ -$\pi$ or $\pi$ -cation interactions in orange, and halogen bonds in green.

Remarkably, all SBS-binding ligands interacted with one catalytic residue (His41) mentioned in previous literature (Table 3). Among the most energy-favoured hits for SBS, Dorsilurin E presented the best binding energy to ISBS (-11.31 kcal/mol). The same compound showed a binding energy of -9.51 kcal/mol in the NSBS. This compound interacted through a hydrogen bond with residue Gln192 from ISBS from a 2.06 Å distance. Besides, hydrophobic interactions with residues His41, Met49, Met165, Glu166, Leu167, Pro168, and Gln189 were observed, which favour complex stabilization. The estimated Kd to the ISBS was 4.78 nM, whereas in the NSBS it was 100.87 nM, for the best inhibitor.

For DS, Sanggenol O was the single best putative inhibitor with a binding energy of -7.29 kcal/mol and a Kd of 4334.61 nM. This energy value was lower than the binding energies for the top hits in the other two putative binding sites. Sanggenol O presented four hydrogen bonds (with residues Met6, Ser113, Gln127 and Arg298 from 2.12, 3.25, 2.60 and 2.21 Å distances, respectively), along with four hydrophobic interactions (Phe8, Pro9, Thr304 and Phe305) (Table 3). Surprisingly, all DS hits inserted at least one of their aromatic rings into a novel pocket comprised by DS and CS residues Met6, Glu290, Arg298; by residues exclusive to CS: Lys5 and Val303; and other residues not included in any site: Ala7, Phe8, Ser113, Gln127, Phe291, Asp295 (Figure 3). The cavity is also present in other crystal structures of the free MPRO (PDB IDs: 6XB1, 6XHU, 6Y2E, 6M2Q, 6Y84 and 7BRO). Interestingly, the mentioned pocket disappears in 6LU7 (MPRO complexed with inhibitor N3). However, in other MPROs complexed with different inhibitors, the novel pocket is present fully or to some degree (PDB IDs: 6M2N, 6WTT, 6XMK, 6M0K, 6W63, 6XBG).

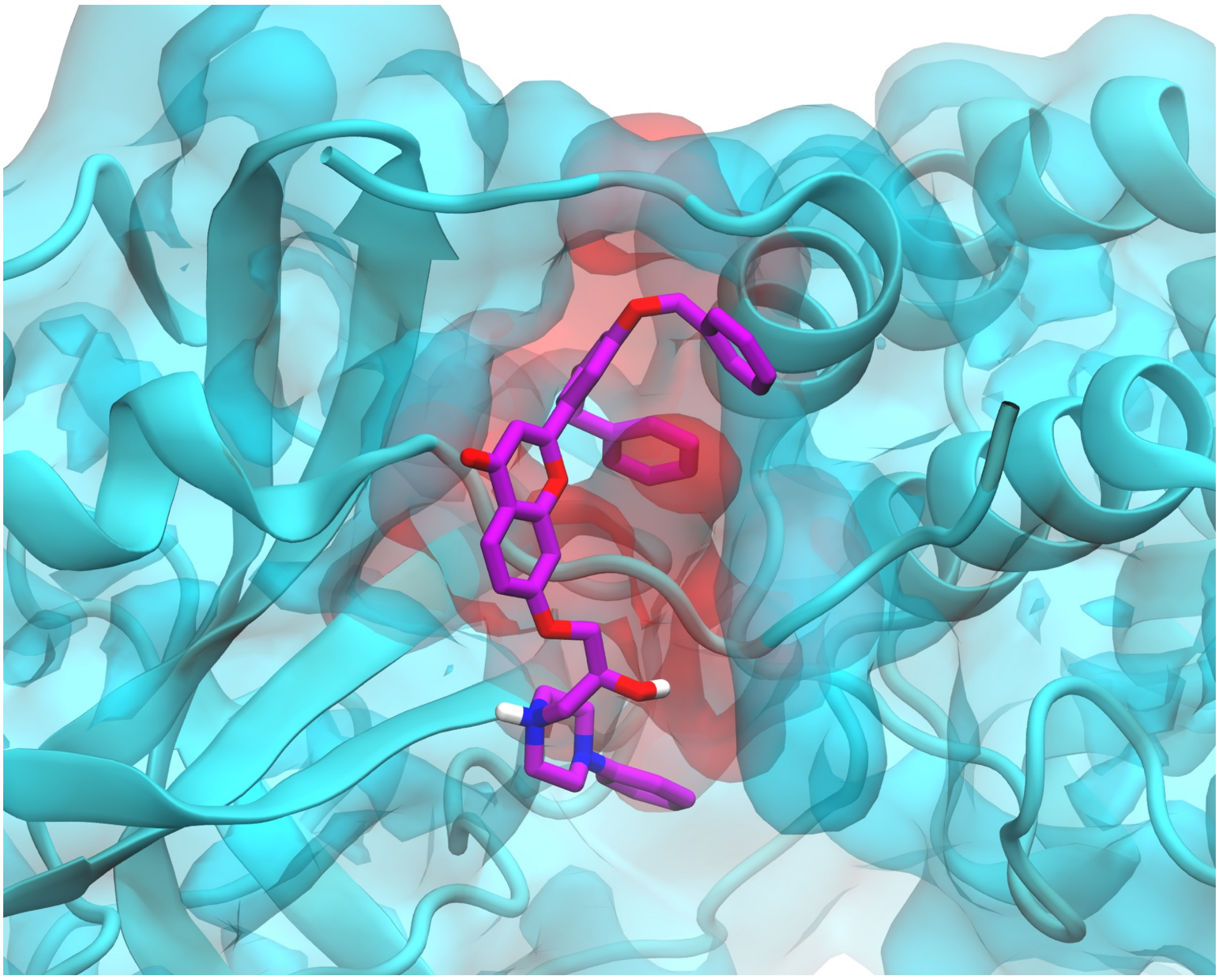

**Figure 3. Close-up of the novel pocket in the dimer binding site.** The surface of residues comprising the novel pocket are shown in red, whereas the rest of the protein's surface is in cyan. A representative compound from those tested against dimerization and cryptic sites, CHEMBL2171598, is shown in magenta sticks as it introduces an aromatic ring into the novel cavity.

**Table 3. Comprehensive list of protein-ligand interactions.**

| Site | Ligand | Torsions | ΔG | Kd (nM) | Hydrophobic | | | Hydrogen bond | | | Other | | | |
|---|---|---|---|---|---|---|---|---|---|---|---|---|---|---|
| | | | | | Residue | Ligand atom | Distance (Å) | Acceptor | Donor | Distance (Å) | Type | Residue | Ligand atoms | Distance |
| Substrate | FL3FQUNP0001 [Dorsilurin E] | 1 | -9.51* -11.31** | 100.87* 4.78** | His41 | C8 | 3.31 | O35 | Gln192 | 2.06 | | | | |
| | | | | | Met49 | C7 | 3.94 | | | | | | | |
| | | | | | Met165 | C3 | 3.47 | | | | | | | |
| | | | | | Glu166 | C21 | 3.17 | | | | | | | |
| | | | | | Leu167 | C29 | 3.99 | | | | | | | |
| | | | | | Pro168 | C27 | 3.85 | | | | | | | |
| | | | | | Gln189 | C3 | 3.70 | | | | | | | |
| | FL2FALNP0014 [Euchrenone a11] | 5 | -9.00* -10.14** | 239.30* 34.70** | His41 | C24 | 3.81 | Glu166 | O35 | 1.94 | | | | |
| | | | | | Met49 | C25 | 3.81 | O36 | Glu166 | 2.02 | | | | |
| | | | | | Phe140 | C30 | 3.85 | | | | | | | |
| | | | | | Leu141 | C14 | 3.41 | | | | | | | |
| | | | | | Met165 | C15 | 3.64 | | | | | | | |
| | | | | | Asp187 | C24 | 3.27 | | | | | | | |
| | FL2FA9NC0016 [Kurziflavolactone C] | 5 | -8.88* -11.24** | 293.25* 5.38** | Thr25 | C15 | 3.82 | Leu141 | O38 | 1.99 | π-Cation | His41 | C8, C9, C10, C11, C12, C13 | 5.83 |
| | | | | | Leu141 | C23 | 3.60 | O39 | Gly143 | 2.36 | | | | |
| | | | | | Asn142 | C31 | 3.79 | Gly143 | O39 | 3.50 | | | | |
| | | | | | Met165 | C3 | 3.55 | O39 | Cys145 | 1.77 | | | | |
| | | | | | Glu166 | C4 | 3.76 | O34 | His163 | 2.23 | | | | |
| | | | | | Val303 | C23 | 3.34 | | | | | | | |
| | | | | | | C24 | 3.07 | | | | | | | |
| | FL2F1AGSN001 [Licorice glycoside E] | 16 | -7.69* -9.26** | 2201.31* 154.06** | Thr25 | C15 | 3.37 | O41 | Thr26 | 1.86 | | | | |
| | | | | | Met165 | C22 | 3.29 | O42 | His41 | 3.08 | | | | |
| | | | | | Pro168 | C32 | 3.48 | O43 | Gly143 | 2.44 | | | | |
| | | | | | Gln189 | C23 | 3.68 | O50 | Ser144 | 3.09 | | | | |
| | | | | | | | | Ser144 | O50 | 2.74 | | | | |
| | | | | | | | | O48 | Ser144 | 2.93 | | | | |
| | | | | | | | | O50 | His 163 | 2.04 | | | | |
| | | | | | | | | O37 | Glu166 | 3.12 | | | | |
| | | | | | | | | O46 | Gln192 | 2.54 | | | | |
| | FL4DACGS0020 [Taxifolin 3'- (6"-phenylacetylglucoside)] | 15 | -7.21* -9.27** | 4963.68* 151.47** | Phe140 | C18 | 3.25 | O40 | Tyr54 | 2.58 | π-Stacking | His41 | C22, C23, C24, C25, C26, C27 | 4.79 |
| | | | | | Met165 | C24 | 3.52 | His164 | O41 | 2.49 | | | | |
| | | | | | Glu166 | C16 | 3.41 | O42 | Glu166 | 1.80 | | | | |
| | | | | | Gln189 | C23 | 3.51 | Glu166 | O42 | 2.13 | | | | |
| | | | | | Gln192 | C4 | 3.77 | Asp187 | O40 | 2.18 | | | | |
| | | | | | | | | Thr190 | O37 | 1.99 | | | | |
| | | | | | | | | O37 | Gln192 | 2.17 | | | | |
| Dimer | FL2FALNP0020 [Sanggenol O] | 3 | -7.29 | 4334.61 | Phe8 | C23 | 3.41 | Met6 | O30 | 44167 | | | | |
| | | | | | Pro9 | C6 | 3.35 | O30 | Ser113 | 3.25 | | | | |
| | | | | | Thr304 | C12 | 3.70 | O30 | Gln127 | 2.60 | | | | |
| | | | | | Phe305 | C3 | 3.48 | | | | | | | |
| | | | | | | C18 | 3.65 | O31 | Arg298 | 2.21 | | | | |
| | CHEMBL2171573 | 14 | -7.28 | 4408.66 | Met6 | C28 | 3.92 | N1 | Ser10 | 3.28 | π-Cation | Arg298 | C25, C26, C27, C28, C29, C31 | 4.81 |
| | | | | | Pro9 | C11 | 3.89 | O42 | Ser10 | 2.35 | | | | |
| | | | | | | C7 | 3.87 | | | | | | | |
| | | | | | Val125 | C15 | 3.16 | Ser10 | O43 | 1.91 | | | | |
| | | | | | Phe291 | C30 | 3.80 | O43 | Gly11 | 2.62 | | | | |
| | | | | | Asp295 | C30 | 3.53 | Glu14 | N1 | 2.61 | | | | |
| | | | | | Gln299 | C37 | 3.34 | | | | | | | |
| | | | | | | C27 | 3.5 | | | | | | | |
| | | | | | Phe305 | C3 | 3.62 | | | | | | | |
| | FL3F1ANP0001 [Kanzonol E] | 4 | -6.92 | 8112.39 | Phe3 | C25 | 3.46 | Met6 | O29 | 2.02 | | | | |
| | | | | | Met6 | C24 | 3.60 | O29 | Gln127 | 3.72 | | | | |
| | | | | | Phe8 | C4 | 3.57 | | | | | | | |
| | | | | | Pro9 | C16 | 3.29 | | | | | | | |
| | | | | | Phe291 | C24 | 3.10 | | | | | | | |
| | | | | | | C25 | 3.14 | | | | | | | |
| | | | | | Asp295 | C22 | 3.07 | | | | | | | |
| | | | | | Arg298 | C6 | 3.98 | | | | | | | |
| | | | | | Gln299 | C25 | 3.63 | | | | | | | |
| | | | | | Thr304 | C12 | 3.45 | | | | | | | |
| | | | | | Phe305 | C10 | 3.92 | | | | | | | |
| | | | | | | C15 | 3.13 | | | | | | | |
| | CHEMBL2171584 | 14 | -6.89 | 8535.30 | Pro9 | C10 | 3.67 | Ser10 | N1 | 1.76 | π-Cation | Arg298 | C18, C19, C20, C21, C22, C23 | 4.74 |
| | | | | | | C11 | 3.87 | | | | | | | |
| | | | | | Glu14 | C35 | 3.92 | O43 | Ser10 | 2.79 | | | | |
| | | | | | Leu115 | C37 | 3.27 | Ser10 | O43 | 2.47 | | | | |
| | | | | | Pro122 | C35 | 3.42 | O42 | Ser10 | 2.58 | | | | |
| | | | | | Val125 | C3 | 3.97 | | | | | | | |
| | | | | | Arg298 | C20 | 3.87 | | | | | | | |
| | | | | | Gln299 | C28 | 3.46 | | | | | | | |
| | | | | | | C20 | 3.39 | | | | | | | |
| | | | | | Phe305 | C14 | 3.49 | | | | | | | |
| | FL2FACNP0014 [Abyssinoflavanone VI] | 4 | -6.83 | 9448.42 | Pro9 | C7 | 3.66 | Met6 | O30 | 2.04 | π-Stacking | Phe8 | C18, C19, C20, C21, C22, C23 | 4.18 |
| | | | | | Thr304 | C8 | 3.76 | O30 | Ser113 | 3.23 | | | | |
| | | | | | Phe305 | C25 | 3.68 | O30 | Gln127 | 2.63 | | | | |
| | | | | | | | | O31 | Arg298 | 2.18 | | | | |
| Cryptic | **CHEMBL2171598** | 14 | -10.59 | 16.19 | Lys5 | C41 | 3.48 | Lys5 | O50 | 2.01 | | | | |
| | | | | | Met6 | C21 | 3.72 | O49 | Ala7 | 2.97 | | | | |
| | | | | | Pro9 | C4 | 3.14 | Val125 | N1 | 44137 | | | | |
| | | | | | Tyr126 | C44 | 3.88 | | | | | | | |
| | | | | | Asp295 | C22 | 3.54 | | | | | | | |
| | | | | | Gln299 | C21<br>C30 | 3.63<br>3.26 | | | | | | | |
| | | | | | Phe305 | C15 | 3.55 | | | | | | | |
| | **CHEMBL2171577** | 14 | -10.51 | 18.54 | Tyr237 | C32<br>C37 | 3.93<br>3.96 | O41 | Lys5 | 1.73 | **Halogen** | Asp197 | Donor: Cl46 | 3.29 |
| | | | | | Asn238 | C35 | 3.70 | O38 | Thr199 | 2.26 | | | | |
| | | | | | Tyr239 | C33 | 3.78 | O42 | Leu287 | 2.18 | | | | |
| | | | | | Leu286 | C7<br>C21<br>C23 | 3.47<br>3.21<br>3.14 | Glu288 | N1 | 2.92 | | | | |
| | | | | | Leu287 | C7 | 3.95 | Glu288 | O41 | 2.02 | | | | |
| | | | | | | | | O40 | Asp289 | 2.44 | | | | |
| | **FL5FAANR0001 [Denticulaflavonol]** | 10 | -10.44 | 20.87 | Lys137 | C7 | 3.69 | O40 | Lys5 | 1.97 | **π-Cation** | Arg131 | C4, C5, C6, C7, C8, C9 | 4.03 |
| | | | | | Thr199 | C15 | 3.85 | O41 | Gln127 | 2.28 | | | | |
| | | | | | Tyr237 | C27 | 3.65 | O39 | Lys137 | 1.97 | | | | |
| | | | | | Tyr239 | C28 | 3.21 | Asp197 | O39 | 2.13 | | | | |
| | | | | | Leu272 | C27 | 3.28 | Glu288 | O40 | 1.94 | | | | |
| | | | | | Leu286 | C21 | 3.63 | O37 | Glu288 | 2.69 | | | | |
| | | | | | Leu287 | C22<br>C18 | 3.68<br>3.52 | O37 | Asp289 | 1.82 | | | | |
| | | | | | | | | O38 | Asp289 | 2.05 | | | | |
| | | | | | | | | O40 | Glu290 | 3.47 | | | | |
| | **FL1CA9NC0001 [Kurzichalcolactone ]** | 8 | -10.23 | 29.79 | Phe8 | C29 | 3.82 | Gly2 | O36 | 2.04 | **π-Cation** | Arg298 | C29, C31, C28, C30, C27, C32 | 4.55 |
| | | | | | Ile213 | C22 | 2.95 | Gln299 | O37 | 2.03 | | | | |
| | | | | | Asp295 | C31 | 3.75 | Val303 | O39 | 1.86 | | | | |
| | | | | | Arg298 | C26 | 3.40 | O38 | Val303 | 2.25 | | | | |
| | | | | | Gln299 | C2 | 3.82 | | | | | | | |
| | **CHEMBL2171578** | 14 | -9.97 | 46.28 | Thr199 | C9 | 3.56 | O37 | Arg131 | 2.80 | **Halogen** | Asp197 | Donor: Cl44 | 3.75 |
| | | | | | Tyr237 | C25 | 3.62 | O37 | Thr199 | 2.08 | | | | |
| | | | | | Asn238 | C27 | 3.44 | O42 | Leu287 | 2.15 | | | | |
| | | | | | Tyr239 | C7<br>C22 | 3.66<br>3.54 | Glu288 | N1 | 3.14 | | | | |
| | | | | | Leu286 | C16<br>C3 | 3.70<br>3.67 | Glu288 | O40 | 1.91 | | | | |
| | | | | | Leu287 | C3<br>C5 | 3.61<br>3.57 | | | | | | | |

Top 5 binding energy flavonoids are listed for each putative binding site. Interactions are according to PLIP predictions. ΔG and $K_d$ values for the substrate binding site are marked as * for the normal docking procedure and ** for the induced-fit docking. Note that both ligand and protein can participate as donor or acceptor in hydrogen bonds. Thus, ligands are represented with their atom name and number, and protein residues are portrayed as their name and number.

The buried rings of CHEMBL2171584 and CHEMBL2171573 sustain a cation-π interaction with Arg298 (Figure 2, Table 3). Abyssinoflavanone VI shows a π-π stacking with Phe8. The inserted rings of Abyssinoflavanone VI, Sanggenol O and Kanzonol E are substituted with hydroxyl groups. In both Abyssinoflavanone VI and Sanggenol O, the hydroxyl group in the position 5 of the backbone ring shows hydrogen bonds with Arg298. Likewise, in both flavonoids the hydroxyl group in the position 7 presents three hydrogen bonds with Met6, Ser113 and Gln127 (Figure 2). Furthermore, the hydroxyl group in the position 7 of the Kanzonol E ring establishes hydrogen bonds with Met6 and Gln127.

As for the CS, CHEMBL2171598 was the best inhibitor with a binding energy of -10.59 kcal/mol and a Kd of 16.19 nM. This compound interacted through hydrogen bonds with residues Lys5, Ala7 and Val125 from 2.01, 2.97 and 2.11 Å, respectively (Table 3). Likewise, it held hydrophobic interactions with residues Lys5, Met6, Pro9, Tyr126, Asp295, Gln299 and Phe305. Interestingly, Kurzichalcolactone and CHEMBL2171598 were found to bind within the novel pocket in the same fashion as the DS hits.

The information regarded surface data for each hit is included in Table 4. The widest interface areas for SBS, DS and CS are observed in Licorice glycoside E with 553.205 Å2; CHEMBL2171573 with 484.185 Å2; and CHEMBL2171598 with 536.563 Å2, respectively.

| Table 4. Protein/Ligand interface areas. | | | | | |
|---|---|---|---|---|---|
| **Site** | **Ligand** | **Name** | **Interface Ligand/Protein (Å$^2$)** | | |
| | | | **SBS** | **DS** | **CS** |
| **SBS** | **FL3FQUNP0001** | Dorsilurin E | 446.243 | --- | --- |
| | **FL2FALNP0014** | Euchrenone a11 | 455.602 | --- | --- |
| | **FL2FA9NC0016** | Kurziflavolactone C | 450.332 | --- | --- |
| | **FL2F1AGSN001** | Licorice glycoside E | 553.205 | --- | --- |
| | **FL4DACGS0020** | Taxifolin 3'-(6"-phenylacetylglucoside) | 452.283 | --- | --- |
| **DS** | **FL2FALNP0020** | Sanggenol O | --- | 332.971 | --- |
| | **CHEMBL2171573** | NA | --- | 484.185 | --- |
| | **FL3F1ANP0001** | Kanzonol E | --- | 360.737 | --- |
| | **CHEMBL2171584** | NA | --- | 482.063 | --- |
| | **FL2FACNP0014** | Abyssinoflavanone VI | --- | 323.374 | --- |
| **CS** | **CHEMBL2171598** | NA | --- | --- | 536.563 |
| | **CHEMBL2171577** | NA | --- | --- | 497.366 |
| | **FL5FAANR0001** | Denticulaflavonol | --- | --- | 477.840 |
| | **FL1CA9NC0001** | Kurzichalcolactone | --- | --- | 428.060 |
| | **CHEMBL2171578** | NA | --- | --- | 505.758 |

Measure of the protein/ligand interface areas expressed in Å² for all 15 hit flavonoids with their respective putative binding sites. Standard deviations provided for comparison.

## 3.6. Positive and negative controls

The inhibitory ligand X77 successfully docked against the ISBS of the free M$^{PRO}$ (-10.77 kcal/mol). A structural alignment between the ligands on the docked structure and the original 6W63 crystal revealed a relatively low deviation from the original (RMSD: 1.352 and 1.693 Å, with and without hydrogens, respectively), considering that the docked ligand was not the original inhibitor (N3) from the employed induced conformation (Figure 4). As for the negative control, the best pose according to the 2Dscore had an energy of -5.46 kcal/mol.

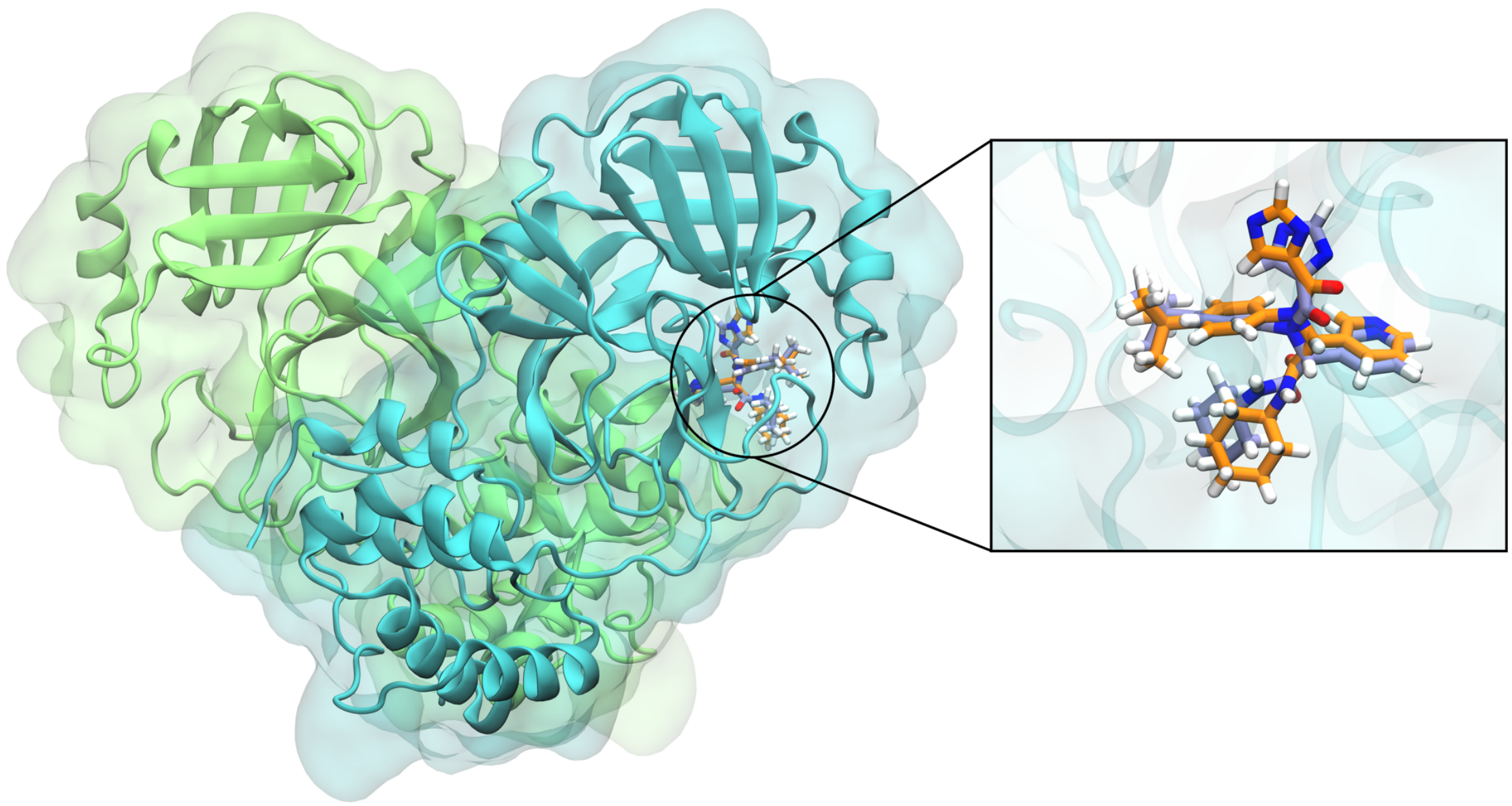

**Figure 4. Structural alignment between the positive control and the original M$^{PRO}$ crystal.** Docked inhibitory ligand was aligned to the original M$^{PRO}$ crystal (PDB: 6W63), revealing an RMSD of 1.352 Å. A close-up is provided such as to observe the docked pose in detail compared to the original position. The M$^{PRO}$ crystal is shown as in Figure 1. The ligand is shown in sticks: orange for the control and ice blue for the cross-docked X77.

## 4. DISCUSSION

### 4.1. Structure

The $M^{PRO}$ binding sites analysis have provided new insights regarding the chemical microenvironments of the structure. Based on the aforementioned pKa's of HSD and HSE, the pH range needed for the coexistence of these two protonation states is 6.7-7.3. Hence, SBS microenvironment's pH is acidic with respect to the pH 7.4 of the general system. Additionally, His172 is proximate to His163, the latter belonging to SBS. His172 is also protonated in the HSD position. Given the proximity of these two 6.5 pKa residues (7.6 Å between alpha carbons), there is more reason to believe in the relative acidity of the SBS. Moreover, the presence of two adjacent histidines with divergent tautomeric states (His163 in HSD and His164 in HSE) suggests said microenvironment can be further divided, being the region around His163 (S1) more acidic than the region around His164 (S2). Said difference is reasonable as His163 is closer to the catalytic subsite S1'. This acidic microenvironment could ease proteolysis as it has been reported that the optimum pH range for catalytic sites in other cysteine-proteases is 5-6 (58–60). Sizeable changes in the microenvironment pH could thus lead to partial or total loss of catalytic function. Therefore, a ligand designed to not just bind the SBS but to change local pH is a hypothetical, but realistic goal for drug designers.

Similar conditions are found in a DS subregion comprising Ser139, Phe140 and Glu166. His163 and His172, close to the aforementioned DS subregion, are in HSD tautomeric state, suggesting an acidic microenvironment for this DS subsection. In contrast, no histidines were observed near or within the CS.

Then, Table 1 shows CS has the highest number of charged residues and thus stands a better chance of establishing saline bonds with an appropriate ligand. The absence of this type of interaction among the other compounds is explained by the electrochemically neutral nature of the hits. Future attempts at drug design might benefit from avoiding electrochemical neutrality for a greater chance of saline bond formation in this region.

### 4.2. Docking

As has already been mentioned, Table 3 contains binding energies which suggest likelihood of effective binding. To take these numbers at face value could be misleading, however. Besides binding energy, binding location and interaction density should also be accounted for. In particular, for the SBS docking, the more subsites a ligand interacts with in a wider area, the higher the chance of proper inhibition, as the cavities needed for the proper accommodation of the natural substrate of $M^{PRO}$ would be sterically obstructed.

For instance, besides a handsome binding energy, Dorsilurin E forms one hydrogen bond and seven hydrophobic interactions with five different SBS subsites (S1', S2, S3, S4 and S5) (Figure 3). By contrast, Euchrenone a11 forms two hydrogen bonds but with the same residue, although hydrophobic interactions are found in fewer subsites (S1', S1, S2 and S3). Similarly, Kurziflavolactone C has five hydrogen bonds, seven hydrophobic interactions (two with the same residue) and a cation-π interaction, but only maintains contact with S1', S1 and S3 besides two residues outside SBS proper. Although they present the least binding energy, Licorice glycoside E and Taxifolin 3'-(6"-phenylacetylglucoside) presented the most hydrogen bonds in the SBS (nine and seven, respectively). Licorice glycoside E interacts thus with S1', S1, S3 and S4 and via hydrophobic interactions with S3, S4 and S5. Besides, it holds a hydrogen bond and a hydrophobic interaction with two separate residues outside SBS proper. Taxifolin 3'-(6"-phenylacetylglucoside) holds hydrogen bonds with S2, S3 and S4; hydrophobic interactions with S1, S3 and S4; and a π-stacking with S1'. Because a wider array of subsites could enhance steric competitiveness, Licorice glycoside E stands a substantial chance at inhibition despite not showing the greatest binding energy.

It is worth noting all the SBS hits show hydrophobic interactions with Glu166, a residue which interacts with the N-terminus domain of the second protomer (56). This intercatenary interaction allows for the stabilization of SBS conformation (in the S1 subsite), which is necessary for $M^{PRO}$ proteolytic activity (61,62). Based on this, it is highly likely that ligands interacting with Glu166 may cause structural modifications on the catalytic site and hinder activity. In addition, π-π stacking interactions were observed between two compounds, Kurziflavolactone C and Taxifolin 3'-(6"-phenylacetylglucoside), and His41, the latter being one of the catalytic residues of SBS. His41 is protonated as HSD. Said tautomer prevails in this region's acidic pH. A slight variation in this microenvironment could change the protonation states and modify the aforementioned interaction. For example, a change in His41 from HSD (single protonation) to HIS+ (double protonation) could intensify bond interaction (63). However, π-π stackings are generally weak. Further insights are needed to assess the real contribution to ligand binding by these interactions.

A last note-worthy point on SBS is that two hits involve glycosylated flavonoids (Licorice glycoside E and Taxifolin 3'-(6"-phenylacetylglucoside)). A previous study on SARS-CoV-1 $3CL^{PRO}$ reported two glycosylated flavonoids as inhibitors, whose sugar moieties interacted with the active sites's S1 and S2 subsites through hydrogen bonds (31). Nonetheless, only Licorice glycoside E's sugar moieties exhibited any interactions whatsoever: hydrogen bonds with His41 (S1'), Gly143 (S1'), Ser144 (S1'), His 163 (S1) and Gly166 (S3). While the visualization tools suggested a non-interactional occupation of S3 by Taxifolin 3'-(6"-phenylacetylglucoside), there is stark contrast with the results reported by the aforementioned study. The differences in sugar-receptiveness between analogous subsites could be illuminating for future studies aiming at deeper comparative analyses between both proteases.

Further along the line, the $M^{PRO}$ DS is presented as a remarkably druggable site for possible inhibition, since a close correlation between the dimerization process and the proteolytic activity has been indicated (64). Remarkably, all the DS hits interacted with highly conserved residues among coronaviruses, namely Arg4, Met6, Gly11, Gly14, Ser10, and Arg298 (32). Under this view, hydrogen bonds might interfere with the dimerization and might be key for future coronavirus inhibitors development (Figure 2, Table 3). Remarkably, all the DS hits interacted at least one of the DS residues which are highly conserved among coronaviruses, namely Arg4, Met6, Gly11, Gly14, Ser10, and Arg298 (32) (Table 3).

Additionally, the binding modes observed in those compounds revealed a cavity partially spanning both DS and CS showing high affinity for substituted and unsubstituted aromatic rings. According to the ring's substitutions, the pocket would allow for π-π and cation-π interactions as well as hydrogen bonds. Both cation-π interactions and hydrogen bonds involve at least one polar group whereas π-π stackings could be present between exclusively nonpolar groups. As the novel pocket has the possibility to form any of these interactions, it must have affinity for both nonpolar or polar moieties. Therefore, the novel pocket has an amphipathic nature. This is coherent with the other results, as the introduced rings can be either hydrophobic (phenyl rings) or hydrophilic (phenyl rings substituted with hydroxyl groups).

The CS may be visually divided in two regions: the first comprised by residues shared with the novel pocket (Lys5, Met6, Gln290, Arg298 and Val303), and the second conformed by the remaining residues (Pro108, Gly109, Arg131, Trp218, Phe219, Tyr239, Glu240, Leu271, Leu272, Leu287, Glu288, Asp289, Glu290 and Gln299). Five of the 11 novel pocket residues are shared

with the CS, and thus have higher CryptoSite scores than the recommended threshold (28). Although this suggests a cryptic nature for this new cavity, further analysis to confirm the reality of the pocket is wanting. By definition, a cryptic site is one that forms in *holo* structure, but not in the *apo* structure (28). As the novel cavity was also present in the *apo* structure of some crystal structures of the $M^{PRO}$ (PDB IDs: 6XB1, 6XHU, 6Y2E, 6M2Q, 6Y84 and 7BRO it would be necessary to see a dramatic modification of this region upon ligand binding over time. Cavity expansion and ligand engulfing upon binding would count as sufficient evidence to formally propose the cavity as a novel cryptic site. Thus, it would be reasonable to submit the systems of the presented best hits to molecular dynamics simulations. Such a method would allow us to evaluate the systems' mechanics and free energy profiles. In contrast to *apo*-$M^{PRO}$, where a free energy local minimum is expected at the novel pocket region, it is presumed that the hits that interacted with said region are able to induce a higher energy conformation and stabilize the pocket (65).

Unlike the novel pocket, the second CS region forms no observable cavity in the *apo*-$M^{PRO}$. If a cavity is then formed upon ligand binding, this region would fit the definition of cryptic site better than the novel pocket. However, as it is the case with the novel cavity, to see this formation and settle the cryptic nature of this second region, it would be necessary to submit the systems to molecular dynamics.

In other considerations, binding energy is not the only factor to take into account, as interface area between ligand and protein in complex is important as well. In short, more interface area involves less SASA, as the ligand buries more protein area. A greater interface area generally enjoys three advantages. First, the ability to sterically compete with the native substrate for the SBS is enhanced as less SASA is present for the natural substrate after the binding event. This fact is closely related to the previously mentioned occupation of the SBS subsites. Second, as the ligand occupies more protein surface, more interfacial water is dispersed and "reinserted" into the solvent, thus partly offsetting the entropic diminishment from ligand binding and preserving the second law of thermodynamics (66). Another contribution to this compensation is an increase in the number of conformational degrees of freedom of the protein itself (67,68). This is the third advantage: a wider contact surface area for the ligand might make it easier to alter the native enzyme conformation for the $M^{PRO}$ and achieve more degrees of freedom. Said correlation has been observed both in protein-protein complexes (69) and individual subunits (70) as a relation between interface surface and sizable conformational change (71). It is suggestive to imagine a similar relationship with regard to protein-ligand complexes. Therefore, while there is not much difference among the interface areas projected by SBS and CS ligands, DS ligands show greater standard deviation.If interface area is therefore taken as an important variable for inhibition, Licorice glycoside E and CHEMBL2171598 could be most promising for SBS and CS, respectively.

## 5. CLOSING REMARKS

Flavonoids have already been suggested as potential inhibitors of the SARS-CoV $M^{PRO}$, showing a high affinity for their hydrophobic aromatic rings and hydrophilic hydroxyl groups (31). In the present research, a group of 15 flavonoids has been identified as possible SARS-CoV-2 $M^{PRO}$ inhibitors. Said group shares some structural similarities with those proposed as SARS-CoV-1 inhibitors. Besides, these compounds have been evaluated against different binding sites in $M^{PRO}$ (SBS, DBS and CBS), and the resulting binding energies were favourable for flavonoid-protein complexes.

Molecular dynamics of the 15 compounds with their respective putative binding sites are currently under way. Moreover, the complete, dimeric structure of $M^{PRO}$ is being considered for SBS as this site is stabilized by both protomers (61,62). Nonetheless, experimental validations are still required to vouch for the effectiveness of the presented compounds as true SARS-CoV-2 inhibitors. Not only PAINS, but colloidal aggregates could induce false positives (53,54), calling for the application of surface tension reduction agents such as Triton-X100 or Tween-80 (72,73).

## 6. ACKNOWLEDGEMENTS

We thank Dr. Miguel Quiliano for sharing his pearls of wisdom with us during the course of this research. We would also like to show our gratitude to the Universidad Privada San Juan Bautista and Mr. Guido Choque for their logistical support.